\pdfoutput=1

\documentclass[10pt]{article}

\usepackage{amsmath}
\usepackage{amssymb}

\usepackage{graphicx}

\usepackage{cite}

\usepackage{color} 

\usepackage[labelfont=bf,labelsep=period,justification=raggedright]{caption}

\makeatletter
\renewcommand{\@biblabel}[1]{\quad#1.}
\makeatother

\date{}

\graphicspath{{../figures/}}

\begin{document}


\begin{flushleft}
{\Large
\textbf{Mapping Dynamic Histone Acetylation Patterns to Gene Expression in Nanog-depleted Murine Embryonic Stem Cells}
}

\medskip

Florian Markowetz$^{1,2,\ast}$, 
Klaas W Mulder$^{1}$,
Edoardo M Airoldi$^{3}$, 
Ihor R Lemischka$^{4}$, 
Olga G Troyanskaya$^{5,\ast}$

\medskip

\textbf{1} Cancer Research UK Cambridge Research Institute, Li Ka Shing Centre, Robinson Way, Cambridge CB2 0RE, UK
\\
\textbf{2} Department of Oncology, University of Cambridge, UK
\\
\textbf{3} Department of Statistics, Harvard University, 1 Oxford Street, Cambridge, MA 02128, USA
\\
\textbf{4} Department of Gene and Cell Medicine and The Black Family Stem Cell Institute, Mount Sinai School of Medicine, New York, NY 10029, USA
\\
\textbf{5} Lewis-Sigler Institute for Integrative Genomics and Department of Computer Science, Princeton University, Princeton, NJ 08544, USA
\\

$\ast$ E-mail: ogt@cs.princeton.edu; florian.markowetz@cancer.org.uk
\end{flushleft}

\section*{Abstract}

Embryonic stem cells (ESC) have the
potential to self-renew indefinitely and to differentiate into any
of the three germ layers. The molecular mechanisms for self-renewal,
maintenance of pluripotency and lineage specification are poorly
understood, but recent results point to a key role for epigenetic
mechanisms. In this study, we focus on quantifying the impact of histone 3
acetylation (H3K9,14ac) on gene expression in murine embryonic stem
cells. We analyze genome-wide histone acetylation patterns and gene
expression profiles measured over the first five days of cell
differentiation triggered by silencing Nanog, a key transcription
factor in ESC regulation. 
We explore the temporal and spatial dynamics of histone acetylation
data and its correlation with gene expression using supervised and unsupervised statistical models. 
On a genome-wide scale, changes in acetylation are
significantly correlated to changes in mRNA expression and, surprisingly, this
coherence increases over time. 
We quantify the predictive power of histone acetylation for gene expression changes in a balanced cross-validation procedure.
In an in-depth  study we focus on genes central to the regulatory network of Mouse ESC, including those identified in a recent genome-wide RNAi screen and in the PluriNet, a computationally derived stem cell signature.
We find that compared to the rest of the genome, ESC-specific genes show
significantly more acetylation signal and a much stronger decrease in acetylation over time, which is often not reflected in an concordant expression change.
These results shed light on the complexity of the relationship between histone acetylation and gene expression and are a step forward to dissect the multilayer regulatory mechanisms that determine stem cell fate.

\section*{Author Summary}

Stem cell differentiation and the maintenance of self-renewal are intrinsically complex processes that require coordinated regulation on many different cellular levels. 
Here we focus on the relationship between two important layers and follow it over the first five days of differentiation. 
The first layer -- measured by acetylation of one of the histone proteins -- describes which parts of the DNA are tightly wrapped up and which lie open. 
The second layer describes the activity of genes measured by their mRNA expression. 
Using a wide array of statistical approaches we show that changes in  histone acetylation are very predictive for gene expression and that the concordance between the two levels increases over time. 
Concentrating on genes central to the regulatory networks in embryonic stem cells we find that key genes show very high acetylation signal in the beginning that decreases quickly over time, indicating that they lie in initially open regions that are rapidly closing down. 
These results are a step forward to a better understanding of the complexities of the relationship between histone acetylation and gene expression, which will help to dissect the multilayer regulatory mechanisms that determine stem cell fate.

\section*{Introduction}

Embryonic stem cells (ESC) are pluripotent cells that have the
potential to self-renew indefinitely and to differentiate into any
of the three germ layers. 
Molecular regulation of embryonic stem cell fate is implemented by a coordinated interaction between epigenetic \cite{Bernstein2006,Boyer2006,Mikkelsen2007,Meissner2008,Vastenhouw2010}, transcriptional \cite{Boyer2005,Loh2006,Ivanova2006,Kim2008,Chickarmane2008,Ying2008} and translational \cite{Sampath2008,Chang2008} mechanisms.

The molecular mechanisms for self-renewal,
maintenance of pluripotency and lineage specification are poorly
understood \cite{Macarthur2009}, but recent results point to key roles for a network of
transcription factors \cite{Wang2006,Kim2008,Nishiyama2009} and a wide range of
epigenetic mechanisms \cite{Szutorisz2005,Boyer2006,Reik2007,Hemberger2009}.
For example, recent work showed the importance of chromatin remodeling factors like polycomb proteins \cite{Boyer2006a,Lee2006} and the SWI/SNF complex \cite{Schaniel2008} for ES cell regulation.
ES cells are richer in less compact euchromatin and, as differentiation progresses, accumulate highly condensed, transcriptionaly inactive heterochromatin regions \cite{Meshorer2006}. 
Major architectural chromatin proteins are hyper-dynamic and bind loosely to chromatin in ES cells. 
Upon differentiation, the hyperdynamic proteins become immobilized on chromatin \cite{Meshorer2006a}. 
Bivalent domains -- consisting of large regions of H3 lysine 27 methylation harboring smaller regions of H3 lysine 4 methylation-- silence developmental genes in ES cells while keeping them poised for action \cite{Bernstein2006,Mikkelsen2007}. 

\paragraph{Multi-layered time-course data in Nanog-depleted mouse ESC}
The number of data sets in ESC linking epigenetic mechanisms to other molecular regulatory mechanisms and following that relationship over time is very limited.
Recently, however, Lu and coworkers \cite{Lu2009} presented a dynamic systems-level study to assess how different molecular regulatory mechanisms interact in stem cell fate decisions in mouse ESC.  Lu \emph{et al}  initiated cell differentiation by experimentally down-regulating Nanog, a key pluripotency regulator. Over the following five days they  measured changes on four different molecular levels: histone acetylation (H3K9,14ac), chromatin-bound RNA polymerase II, messenger RNA (mRNA) expression and nuclear protein abundance. This data set provides a rich resource to untangle the complexity of the multi-layer regulatory mechanism responsible for stem cell fate. Lu \emph{et al} anchored their analyses on changes in nuclear protein expression and found that many lacked concordant changes in mRNA expression, pointing to important roles for translational and post-translational regulation of ESC fate. Here, we complement theses analyses with an in-depth study of the relation between histone acetylation and gene expression in the same data set.

\paragraph{Histone acetylation and gene expression}
The acetylation of
lysine residues is among the best characterized histone modifications. 
It has long been correlated with transcriptional
activation \cite{Allfrey1964,Pogo1966}. This observation has been
verified in many recent high-throughput studies
\cite{Sterner2000,Kurdistani2003,Karlic2010}. For example, histone
acetylation was found to be positively correlated with expression in
yeast \cite{Kurdistani2004,Pokholok2005} and human T cells
\cite{Wang2008,Roh2005}. 
The last study also suggests that acetylation sites often cluster together in so called `acetylation islands' \cite{Roh2005}.

Several models have been suggested to explain how histone acetylation and other modifications regulate gene expression \cite{Schones2008}, including charge neutralization \cite{Shogren-Knaak2006} and a signalling pathway model \cite{Schreiber2002}. 
However, the detailed mechanism is still poorly understood. 
This problem is highlighted by two recent studies, one experimental and one statistical. 
G{\"u}nther \emph{et al.} \cite{Guenther2007}  stress the importance of additional regulatory events by showing that acetylated and methylated nucleosomes, as well as RNA polymerase II, occupy the promoters of most protein-coding genes in human ES cells, even those that are not expressed.  
Yuan \emph{et al.} \cite{Yuan2006} assessed the global regulatory role of histone acetylation in \emph{Saccharomyces cerevisiae} by controlling for confounding effects like transcription factor binding sites and nucleosome occupancy. 
They find a clear effect of histone 3 acetylation, but no significant direct impact of histone 4 acetylation or combinatorial effects, even though they correlate with expression.

These results indicate that further experimental results and statistical
analyses are required to untangle the regulatory role of histone
acetylation and the mechanism by which it acts. 
The need for a better understanding of histone acetylation is especially urgent in ES cells, where many key regulatory mechanisms are epigenetic and act by chromatin modifications and remodeling. 
For example, embryonic stem cells in which histone de-acetylation is inhibited, undergo morphological and gene expression changes indicative of differentiation \cite{Karantzali2008}.

\paragraph{Overview of results}
In the following, we first start by analyzing the internal structure of the histone acetylation profiles and their change during differentiation. 
We investigate the dynamics of acetylation over time and find that the location of acetylation islands remain stable. 
We find  that differentially down-regulated genes are accompanied by a much stronger loss of acetylation than up-regulated genes are by a gain of acetylation. 
In a next step we assess the dynamics of the correlation between mean acetylation levels and expression and find that coordination increases over time. 
Using statistical classification methods we then quantify the predictive power of acetylation profiles for gene expression changes.
Finally, we focus on genes playing key roles in the regulatory networks governing fate decisions in embryonic stem cells.
We show that these genes show highly increased acetylation profiles. 
Over time the high levels of acetylation get reduced more strongly than in other, not ESC specific genes. 
This behaviour is far less pronounced in the gene expression data, pointing to a key role in non-transcriptional regulation of pluripotency for important ESC genes.

\section*{Results}

Our central questions are how changes in gene expression are reflected in histone acetylation, how predictive histone acetylation is for gene expression changes, and how this relationship changes over time.
To answer them, in the following we employ different statistical approaches to describe the internal structure of histone acetylation profiles and to map them to changes in gene expression.

\subsection*{Histone acetylation changes in differentially expressed genes}

\subsubsection*{Location of acetylation islands is stable over time}

As examples of the data we work with, Figure~1A shows acetylation profiles of Pou5f1/Oct4 and Klf4. The plots show acetylation levels  at four time points: before silencing Nanog (day 0) and at days 1, 3, and 5 afterwards. 
The plots show large internal variation of acetylation signal for each gene.
As a first preparatory step in our analysis we investigated if there is evidence that the location of acetylation signal changes over time.
If the signal location does not change, then only the quantitative level of acetylation are important when mapping it to gene expression in the next steps of our study.

We identified acetylation islands \cite{Roh2005} by comparing probe signal to background distribution of control probes on the array (see \emph{Material and Methods}). 
Figure~1A depicts the background distribution as a grey area, all probes above it are counted as `acetylated', all probes inside as `unacetylated'. (This is a slight abuse of terminology since technically it is not the probe that is acetylated but the histone protein bound to a piece of DNA complementary to the probe.)
With these results, we investigated dynamical changes on the probe level and asked for each gene:  Are the same probes acetylated over time, or does the position of acetylation signal change over time? 
To answer this question, we represented each gene by two numbers: the percentage of probes staying un-acetylated and the percentage of probes staying acetylated between time-points. Figure~1B shows that the distribution of these values is concentrated in the upper right corner of the plot which corresponds to perfect conservation of acetylation location over time. 
Probes that are acetylated at any time-point stay acetylated and un-acetylated probes stay un-acetylated. 
In a second step we investigated if regions of peak signal in the binned profiles change over time. 
We defined a peak as those bins that include the maximum of the profile and together carry $\geq30\%$ of the signal. 
Figure~1C shows that not only acetylated probes but also peaks stay stable over time.

Thus, in summary, in our data we find no evidence that the location of acetylation signal changes over time. 
This simple analysis plays only a preparatory role in our study: it allows us to focus on quantitative changes in signal intensity in the next steps of our analysis.
The data we work with from now is exemplified by the blue heatmaps underneath the profiles in Figure~1A.
For each gene, our data captures the quantitative acetylation signal in a region of $\pm3.5$kb around the transcription start site (TSS).

\subsubsection*{Loss of acetylation is more pronounced than gain of acetylation}

We find clear correlations between histone acetylation and gene expression. 
For example, Figure~2A shows all genes differentially expressed on day 5. 
In this plot, genes transcriptionally up-regulated also show increased levels of acetylation, while down-regulated genes show a decrease. 

However, these plots also indicate that the loss of acetylation for down-regulated genes is much stronger pronounced than the gain of acetylation for up-regulated genes. 
This is particularly visible in Figure~2B, which plots the acetylation distributions separately for up-regulated, down-regulated and stable genes. 
The down-regulated genes show a very strong loss over the whole width of the profile, while the up-regulated genes show a much weaker signal and only close to the TSS.
Genes without significant expression changes show a strong bias towards loss of acetylation, but the size of the effect is much smaller than in the down-regulated genes.

\subsubsection*{Partial correlation analysis resolves spatial and temporal dependencies in acetylation profiles}

We were interested in the internal correlation structure of the histone acetylation profiles and used partial correlation analysis (see \emph{Material and Methods}) to measure the direct relations between regions around TSS (i.e. the bins in the profile).
Figure~3A shows partial correlation matrices combining data from day 1, 3 and 5.
We computed one matrix for genes differentially expressed on day 5 and a second one for genes with stable expression.
Both matrices show a strong stripe-pattern indicating high correlation for neighboring bins and for the same bin at different days.
The differences between the two matrices are minimal, as can be seen in the right-most matrix of Figure~3A. 
Only between day 3 and 5 do the acetylation profiles in differential genes show a little bit more correlation than those in the non-differential genes.

The significant entries of the partial correlation matrix can be depicted by the graph structure shown in Figure~3B.
The three layers of the graph correspond to the three days and each edge indicates a significant partial correlation coefficient.
We show the graph for the non-differential genes since their larger number results in higher power.
The graph shows that the spatial and temporal dependencies between variables very clearly show in the correlation structure of the data, for example almost all neighboring bins at the same time-point are connected.
However, close to the TSS the graph is much less connected than in more distant regions.
This possible reflects the presence of nucleosome free regions around the TSS in many active genes \cite{Yuan2005}.

\subsection*{Coordination of histone acetylation and gene expression increases over time}

We assessed the correlation between histone acetylation changes and gene expression changes versus day 0 for all pairings of days. 
This analysis is anchored on the ESC state (day 0) and assesses the coordination of {\em cumulative} changes away from it.
In a first step, we summarized each acetylation profile by the mean and computed the standard Pearson correlation between the resulting acetylation vector and expression vector (Figure~4A, left matrix). 
The results show that changes in acetylation demonstrate significant correlation to changes in mRNA expression. 
Correlations with acetylation changes on day 1 are generally small (in general $<0.1$), but correlations between changes on days 3 and 5 show very significant values, e.g. on day 5 Pearson correlation is 0.344. 
Even though this value is small, the level of coherence is very surprising given the large number of genes ($>17\ 000$). 
The correlation table shows coherence between histone acetylation and gene expression increases over time and is biggest on day 5.

\subsubsection*{Correlation results are statistically significant}
We assessed the significance of observed correlations in two ways.
First, we used the analytic Null-distributions known for the correlation measures we used \cite{Anderson2004}. 
Significance is a function of sample size and with $>17\ 000$ genes we find all correlations between days 3 and 5 to be significant with $p$-values smaller than $10^{-100}$. 
Correlations with day 1 (first row or column in Figure~4A left matrix) are much weaker, but still almost always significant on a level of $10^{-4}$. 
One reason for these extremely small $p$-values is that the analytic Null-distributions assume independence between genes, which is an unreasonable assumption for genomic data. 
To correct for this bias, we used a permutation approach that keeps the correlation structure of genes intact for a second assessment of significance. 
We compared the correlations measured in the actual data with the distribution of $10^4$ correlation values computed on permuted versions of the data.
However, qualitatively the results were identical to the first approach: correlations between days 3 and 5 are very significant (no permutation yielded a correlation exceeding the value on the actual data) and correlations to day 1 are much weaker.

\subsubsection*{Correlation results are robust to gene selection and correlation measures}
In the next step we assessed the robustness of the observed correlation pattern by using different types of correlation measures, different ways to average the acetylation changes and different subsets of the data (right matrix of Figure~4A). 
In particular, we used the Spearman rank correlation between the \emph{median} (instead of \emph{mean}) acetylation change and expression, as well as the correlations computed by Canonical Correlation Analysis, a statistical method to find directions of maximal correlation between datasets (see \emph{Materials and Methods}). 
For each of these different ways to compute correlations, we asked whether the results are global or driven by a small subset of genes, e.g. the differentially expressed genes. 
Figure~4A summarizes our finding that the pattern of increased correlation over time was preserved for all subsets of genes and definitions of correlation.
This indicates that our results are reproducible and describe a global event not limited to a specific subset of genes or a particular correlation estimate. 

\subsection*{Acetylation changes are highly predictive for gene expression changes}

Correlation analysis showed global coherence between \emph{averaged} acetylation profiles and gene expression.
Next, we analyzed the predictive power of the \emph{complete} profile using a wide array of statistical classification methods.
We investigate the predictive power of histone acetylation for gene expression by asking: 
Can changes in histone acetylation patterns project changes in gene expression? 
If acetylation is a marker for open chromatin, does it predict expression change in general,  and how well can it distinguish up- from down-regulation?

\subsubsection*{Setup of classification analysis} 
To address these questions we applied a comprehensive collection of classification methods in an unbiased repeated 10-fold cross-validation study (see \emph{Material and Methods}) to four different classification problems:  
(i) Distinguishing transcriptionally up- from down-regulated genes, 
(ii) distinguishing down-regulated genes from un-responsive genes, 
(iii) distinguishing up-regulated genes from un-responsive genes, or 
(iv) distinguishing differential genes (up or down) from un-responsive genes. 
On each of these four problems we used 
(a.)~Support Vector Machines with different kernel functions; 
(b.)~versions of Gaussian discriminant analysis; 
(c.)~several classification tree methods; 
(d.)~$k$-nearest neighbor classification with varying numbers of neighbors; as well as 
(e.)~naive Bayes classification, neural networks and logistic regression (see \emph{Material and Methods}). 

Different classifiers may respond to different signal in the data. 
For example, naive Bayes classifiers assume independence of features (here: the bins in the acetylation profiles), while SVM and other non-linear classifiers can make use of interactions between features.
Our selection of classification methods offers a comprehensive overview of current state-of-the-art methodology and makes our results independent of an arbitrary choice of some particular classification method. 

\subsubsection*{Results of classification analysis}
Figure~4B shows the results of the cross validation study.
In all problems all classifiers clearly beat the baseline of 50\% accuracy, but there are obvious differences in performance: Distinguishing up- from down-regulated genes is the easiest problem with performances reaching 80\% and above.
This margin of improvement over baseline is quite large given that predicting expression from sequence information is a notoriously hard problem (see the discussion of \cite{Beer2004} in \cite{Yuan2007}) and that the acetylation marks we are using ranked  far behind others in  predictive power for expression in a recent comparison \cite{Karlic2010}.

The other three classification problems are harder, especially for distinguishing differential from unresponsive genes classifier performances only reach a level of around 60\% accuracy. 
This can be explained by the set of differential genes containing two opposing signals, which makes it hard to clearly separate it. 

The other two curves in Figure~4B show that down-regulated genes can be better distinguished from un-responsive genes than up-regulated genes can. 
This might be surprising since we saw in Figure~2B that the acetylation  profile distributions for down-regulated genes overlapped more with the un-responsive genes than the profiles for up-regulated genes did. 
However, it can be explained by the fact that loss of acetylation affects wider regions than gain of acetylation signal as can be seen in Figure~2B. 

For all classification problems, more highly regularized and constrained methods beat less regularized ones; for example, a larger  number of neighbors  improves k-nearest neighbor classification, quadratic Gaussian discriminant analysis performs worse than the three linear versions, and the higher degree polynomial SVMs are in most cases out-performed by the linear SVM.

\subsection*{ESC genes show very strong acetylation changes, which are not all reflected in gene expression}

Our results so far investigated the general relationship between histone acetylation and gene expression. 
Now we focus on sets of genes central to the regulatory network governing ES cell state.
We will call them \emph{ESC genes} for short.
We used several  freely available data sources, which complement each other in describing ESC from different perspectives including transcriptional, proteomic and functional. 
In particular, we used five different descriptions of key ESC genes given by
(1) the PluriNet \cite{Mueller2008}, a computationally derived stem cell signature; 
(2) hits of a recent RNAi screen for self-renewal \cite{Hu2009}; 
(3) gene ontology \cite{Ashburner2000} term GO:0019827 `stem cell maintenance'; 
(4) members of an ESC-specific protein-interaction network \cite{Wang2006};  
(5) key transcriptional regulators of ESC \cite{Ivanova2006}. 

\subsubsection*{Average histone acetylation signal is very high in ESC genes}

The left panel in Figure~5A plots the sorted mean acetylation signal on day 0, before Nanog-silencing triggers differentiation, and underneath the ranks where the five ESC gene lists fall in this ordering. 
In all five gene lists we observe a strong trend for ESC genes  to have a very high average acetylation signal, i.e. the bars representing the gene sets all cluster on the right-hand side of the plot.
The trends are strong and easily visible by eye; we quantify their significance by Gene Set Enrichment Analysis (GSEA \cite{Subramanian2005}, see \emph{Materials and Methods}) and observe $p$-values $\leq10^{-4}$  for four  gene sets and $p < 10^{-3}$ for the fifth one.

\subsubsection*{Decrease in histone acetylation signal is not accompanied by similarly strong decrease in gene expression}

Over time the acetylation signal generally diminishes, but this trend is especially pronounced in ESC genes (Figure~5A, middle panel). 
All five gene sets have $p$-values $\leq10^{-2}$ and three of them even $\leq10^{-4}$.
This shows that compared to all other genes, ESC genes are predominantly affected by de-acetylation during the first days of differentiation.
If we take histone acetylation as a marker of open chromatin, this result could indicate that the chromatin regions, at which the ESC genes are located, are closing down over time.

We were then interested in seeing how this strong de-acetylation is reflected in gene expression (right-most panel in Figure~5A).
Qualitatively, the correlation results of Fig.~4A also hold for the sets of ESC genes.
However, when comparing expression changes in ESC genes to other genes, we only found a strong trend to negative expression changes in the set of transcriptional regulators \cite{Ivanova2006} ($p\leq10^{-4}$) but only much less in the other gene sets.
Members of the protein interaction network \cite{Wang2006} show moderate down-regulation, but in particular the PluriNet genes \cite{Mueller2008} and the RNAi hits \cite{Hu2009} are uniformly spread out over the spectrum.
One way to interpret this observation are other major regulatory influences on key ESC genes that can not be explained by accumulation of condensed and transcriptionally inactive heterochromatin regions (as far as these are indicted by histone de-acetylation).

\subsubsection*{ESC genes show overall very strong histone acetylation profiles}
The high acetylation signal of ESC genes is not only found in the mean value, but over the whole profile. 
Figure~5B compares the distribution of acetylation signal between ESC genes and all other genes for each bin individually. 
Quantiles for the global acetylation distribution across all genes are shown in blue and white boxplots represent the distributions for the union of ESC gene sets described above.
Because of their important regulatory function, the set of transcriptional regulators \cite{Ivanova2006} are additionally highlighted in red.
We see a significant upwards shift for ESC genes in over the whole range of the profile.
This shift is especially pronounced for stem cell transcriptional regulators directly before the TSS.

\section*{Discussion}

In this paper we have addressed several questions central to an understanding of the relationship between histone acetylation and gene expression. 
Using a wide array of methods we have investigated how changes in gene expression are reflected in histone acetylation, how predictive histone acetylation is for gene expression changes, and how this relationship changes over time.
In the following we will give a short discussion of our main results.

\paragraph{Gain and loss of acetylation over time}

While there are less genes transcriptionally down-regulated than up-regulated (Fig.~2A) we find that the accompanying de-acetylation events are much more pronounced than the acetylation events.
The wider impact of de-acetylation could be seen in the heatmap (Fig~2A) and the distribution plots (Fig~2B). 
Its effects could be seen in the results of correlation analysis (Fig~3) and classification analysis (Fig~4).

While Roh \emph{et al} \cite{Roh2005} observe main changes in a region of $\pm1$kb  around TSS, we observe wider changes especially for down-regulated genes. 
In particular for ESC genes we find strong acetylation changes over time (Fig~5A) and for several transcriptional regulators we see that acetylation is extremely high before TSS (Fig~5B). 
These differences in acetylation signal could point to mechanistic differences in how acetyatlion acts and which transcriptional co-factors it recruits in activated and repressed genes.

\paragraph{Predictive power of acetylation changes for gene expression changes}

We have seen  from the classification results (Figure~4B)  that histone acetylation changes are highly  predictive of gene expression changes.
We have also found that the coordination between histone acetylation measurements and gene expression increases over time. 
This pattern is stable to varying correlation measures and selecting subsets of genes (Figure~4A).

One way to interpret this trend is a time-lag before changes in chromatin structure (as far as these are indicated by histone acetylation) result in coordinated changes in gene expression. 
In this scenario, chromatin changes induce gene expression changes, which only become visible at a later time-point and thus increase correlation over time.
However, the time-delay in our case would span several days and it is not clear which mechanism causes it, since (de-)acetylation dynamics --at least in yeast-- are known to work in the order of minutes \cite{Kurdistani2003}.
Another question we can not answer from predictive models alone is whether chromatin structure changes are \emph{causative} for gene expression changes or whether it is the other way round: chromatin changes could be induced by expression changes and activation of chromatin modelling proteins.

\paragraph{Distinct acetylation patterns in key ESC genes}

It is known that ES cells in general are rich in less compact euchromatin \cite{Meshorer2006} and  high histone acetylation levels are one of the indicators for these open chromatin regions. 
Thus, the strong acetylation signal of ESC genes we observed could indicate that they are located at open chromatin and thus easily accessible to transcription factors. 
Our results show that ESC genes are enriched for strong de-aceylation (Figure~5A; middle panel).  
This observation could point to the fact that in early development, as soon as the cell commits to a certain lineage,  ESC are located in genomic regions that are de-acetylated and compacted much faster than other regions of the genome.
Our interpretation depends on how close the link between histone acetylation and chromatin structure actually is.  Not all chromatin changes will be reflected in histone acetylation and in future work it will be important to also probe other markers of chromatin organization, like \emph{e.g.} histone methylation, in ESC over time. Integrated analyses of different markers will give a much richer picture of epi-genetic gene regulation than any individual marker can \cite{Karlic2010}.

The stability of acetylation islands we observe and the strong de-acetylation over time agree with a \emph{global accessibility model} of lineage commitment \cite{Szutorisz2005} in which ES cells are subject to global active histone modifications that get lost in a lineage-specific way during differentiation. In contrast, our observations do not agree with a \emph{localised marking model} \cite{Szutorisz2005} in which short regions of accessible chromatin are expanded during development. This expansion would be visible as location changes in acetylation islands which we did not observe. However, the situation could change if the time-course was repeated using ChIP-seq instead of ChIP-chip technology which offers a higher resolution of acetylation changes.

Our results have two important implications: First, the pattern in Figure~5A shows that the expressions of some of the key ESC genes, especially PluriNet and the RNAi hits, are not regulated completely by chromatin accessibility (as far as it is visible in histone acetylation patterns). 
Second, the uniform distribution of gene expression changes in many ESC genes shows that they do not regulate pluripotency on a transcriptional level.

The differences in behaviour  we see between transcriptional regulators on the one hand and the PluriNet genes and RNAi hits on the other hand could possibly be attributed to differences in how specific these genes actually are for ES cells.
The transcriptional regulators are all well-known and very specific, while the computational and functional predictions from PluriNet or RNAi screens can also capture many non-specific genes. 
For example, the MATISSE algorithm \cite{Ulitsky2007} used to derive the PluriNet signature uses protein-interactions and gene expression to find genes connected to key ESC markers.
The genes `pulled in' by the algorithm can help to better understand the mechanisms behind the known marker genes, without being specific regulators themselves.
Similar considerations hold for RNAi screens.
Many genes contributing to basic cellular functions can potentially be found to be essential for self-renewal, without being stem-cell specific.

\paragraph{In summary}
Our results are a step forward to a better understanding of the complexities of the relationship between histone acetylation and gene expression, which will help to dissect the multilayer regulatory mechanisms that determine stem cell fate.
The data of Lu \emph{et al} \cite{Lu2009} is an example of a very rich and complex dynamic phenotype of a single-gene perturbation. 
Future work will need to integrate this data with similar phenotypes of other genes and then use statistical methods \cite{Markowetz2010} to uncover the cellular networks underlying the observed phenotypes.

\section*{Material and Methods}

  \paragraph{Software}

The complete analysis was performed in the statistical computing language R \cite{R} using packages available from the Bioconductor website at \texttt{http://www.bioconductor.org} \cite{bioconductor}. In addition to the basic distribution we mainly used the packages \texttt{limma} \cite{limma,Ritchie2007}, \texttt{GeneNet} \cite{genenet}, \texttt{CCA} \cite{gonzalez07cca}, \texttt{MLInterfaces} \cite{mlinterfaces}, and all packages implied by these. All code is available from the first author upon request.

  \paragraph{Data preprocessing}

Data generation, pre-processing and mapping of genes between datasets is done in exactly the same way as in \cite{Lu2009}. 
Per day we use for each gene the average of three replicates of gene expression measurements and the average of two replicates of H3K9,14ac ChIP-chip. 
We apply simple quality filters to the histone acetylation data: 19, 413 genes are represented by probes on the chip. 
For each gene, the probes are concentrated in a $\pm3.5$ kb region around transcription start. 
Out of the 19, 413 genes, we select the 17,268 that have more than 10 probes within 3.5kb of transcription start. 
On average, we find $\sim30$ probes per gene, which typically have a distance of $\sim$248 bases pairs. 
For all of these genes the data set also contains gene expression measurements.

  \paragraph{Identification of acetylation islands}

To find acetylation islands \cite{Roh2005}, we compared the measurement for each probe against the distribution of measurements of the control probes on the array. 
The control probes are designed to be un-acetylated and thus constitute a negative control.
Comparing the probe values against the Null distribution yields a $p$-value for every probe. 
We use an FDR cutoff of $\alpha=0.1$ on the $p$-value distribution to decide which probes to call acetylated and which not. 

  \paragraph{Frequency of acetylation changes}

We computed for each gene the conditional distribution of probe
acetylation states given the previous time-point. The distribution table can be represented
by two numbers: the percentage $\alpha$ of probes staying
un-acetylated and the percentage $\beta$ of probes staying
acetylated. In this way, each gene can be mapped to a point in $[0,
100] \times [0, 100]$. Genes with too few ($\leq 3$) acetylated or un-acetylated probes
(=3510 genes) were discarded because their estimates would be
unstable. Results for the remaining 13758 genes are shown in
Figure~2A. Plotted are the frequencies computed by assuming that
the change distribution is the same for all time points; results
don't change qualitatively if we compute individual changes between
days (see inlay in Figure~2A for genes differential on day 5).

  \paragraph{Step-wise linear approximation of acetylation profiles}

Genes are represented by different number of probes with varying
distances between each other and to the transcription start site. To make acetylation profiles
comparable between genes we map them onto vectors of equal length by
averaging all probes in equi-distant bins around transcription
start. We chose a binning of $0.5$kb,
thus covering the $\pm3.5$kb region with 14 bins and mapping each
acetylation profile into $R^{14}$. We only considered the signal
above background, bins with no probes above background were set to
zero. Examples of raw and binned profiles can be seen in Fig.~1A.
This binning and averaging makes the data comparable between genes, while preserving most of the quantitative variation in the data.

  \paragraph{Partial-correlation analysis}

To delineate the correlation structure of the data we used partial
correlation analysis, also called a Gaussian graphical model
\cite{Lauritzen1996,Anderson2004}. In contrast to regular
correlation, \emph{partial} correlation corrects for the influence
of all other variables in the model: Vanishing partial correlation
(under a Gaussian assumption) means that two variables are
independent given all other variables (genomic regions in our case).
Thus, partial correlation coefficients measure the direct
relationship between two variables, while regular correlation
coefficients also measure indirect effects. We used a shrinkage
approach \cite{schaefer05shrinkage} for robust estimation of
partial correlations. The results can be depicted in a graph, where
each node corresponds to a variable (a genomic region) and each edge
a partial correlation that is different from zero. Missing edges
indicate vanishing partial correlation and thus conditional
independence. We select the network containing only edges with
probability $> 0.9$ corresponding to a local FDR cutoff of $0.1$
\cite{genenet}.

  \paragraph{Canonical correlation analysis}
Canonical correlation analysis (CCA, see \cite{Anderson2004}) is a
way of measuring the linear relationship between two
multidimensional variables. In general, CCA finds vectors $a$ and
$b$ such that the random variables $a'X$ and $b'Y$ maximize the
correlation $\rho = \text{cor}(a' X, b' Y)$. Vectors $a$ and $b$ are
unique up to scalar multiplication. The random variables $U = a'X$
and $V = b'Y$ are the first pair of canonical variables and $\rho$
is called the canonical correlation.  In our application $X$ corresponds
to the histone acetylation data (a 14 dimensional random variable)
and $Y$ to the RNA data per day (a one dimensional random variable).
Thus, we only need to find vector $a$ to maximize the correlation
between the two data sets. Computing the correlation between mean
acetylation profiles and expression is closely related to CCA, since
it corresponds to the choice of
$a_{\text{mean}}=\frac{1}{14}(1,\ldots,1)$, but it is not guaranteed
to find the maximal correlation.

  \paragraph{Classification methods}
(a.)~Support Vector Machines (SVM, \cite{Scholkopf2002}) construct
the hyperplane with maximal margin of separating between the
positive and negative training examples. Using non-linear distance
measures, so-called kernel functions, this approach can be extended
to non-linear classification. We use a linear kernel, a radial basis
function kernel and polynomial kernels of degrees 2 and 3.
(b.)~Gaussian Discriminant Analysis \cite{Hastie2001} assumes that
the positive and negative examples follow a multivariate normal
distribution. Versions of Discriminant Analysis differ by the
constraints they put on the covariance matrices: no constraints
(Quadratic DA); or the same covariance matrix for both classes
(Linear DA); or  the same \emph{diagonal} covariance matrix
(Diagonal Linear DA). Stabilized Linear DA is linear discriminant
analysis based on left-spherically distributed linear scores.
(c.)~Classification trees \cite{Breiman1984} recursively partition
the dataset by splitting along most-informative single features.
Bagging \cite{Breiman1996} (short for 'bootstrap aggregating')
aggregates many classification trees built on resampled versions of
the training data. Similar to bagging, a Random Forest
\cite{Breiman2001} is an aggregation of many classification trees
built on resampled versions of the data and on a randomly chosen
subset of features. (d.)~$k$-nearest neighbors predicts a gene into
the class represented by the majority of the $k$ genes closest to
it. We use $k$=1, 5, 10, and 15. (e.)~Naive Bayes classification
assumes independence of features (hence \emph{naive}) and classifies
according to the class posterior probability. The neural network
\cite{Ripley1996} is a single-hidden-layer network. Logistic
Regression \cite{Hastie2001} combines a linear model of the data
together with a logistic function to model class probabilities. All
classifiers were used via the R-package \texttt{MLInterfaces}
\cite{mlinterfaces} and with the default parameters defined there.

  \paragraph{Balanced evaluation of prediction accuracy}

The datasets we use for classification can be very unbalanced, for example only $\sim$5\% of all genes show a
significant expression change. Thus, the baseline for classification
is already at 95\% accuracy (when we predict all genes as
'unchanged'). To be able to compare between methods and different classification scenarios, we resorted to a random sampling strategy: We sampled from the larger part of the training set 20 times sets of the size of the smaller part. This
created 20 instances of balanced training sets with a
baseline of 50\%. On each training set we computed the 10-fold
cross-validation (CV) accuracy. The variance we see in the CV results is thus a sum of the variance introduced by sampling the training set and the variance from randomly splitting the data into 10 subsets inside CV procedure. It is reassuring that  Fig~4B overall shows very consistent results, only individual boxplots are spread out widely.

  \paragraph{Gene Set Enrichment Analysis (GSEA)}
  
The goal of GSEA \cite{Subramanian2005} is to determine whether
members of a gene set (for example ES genes) tend to occur toward
the top (or bottom) of a list of phenotypes (in our case: mean
acetylation or expression). GSEA is
especially suited to find coherent changes in a group of genes, even
if the individual changes are small.
GSEA calculates an enrichment score for
a given gene set using rank of genes and infers statistical
significance of each ES against ES background distribution
calculated by permutation of the original data set. We report the empirical $p$-value after  $2\cdot10^4$ permutations,
\emph{i.e.} in how many permutations did we observe a result more
extreme than the one on real data. We did no multiple-testing correction, since with only 15 tests altogether even the most conservative correction ($p' = 15 \cdot p$) would not qualitatively change our results.

\section*{Acknowledgments}


\newpage
\section*{Figures and  legends}


\begin{figure}[h!]
\includegraphics{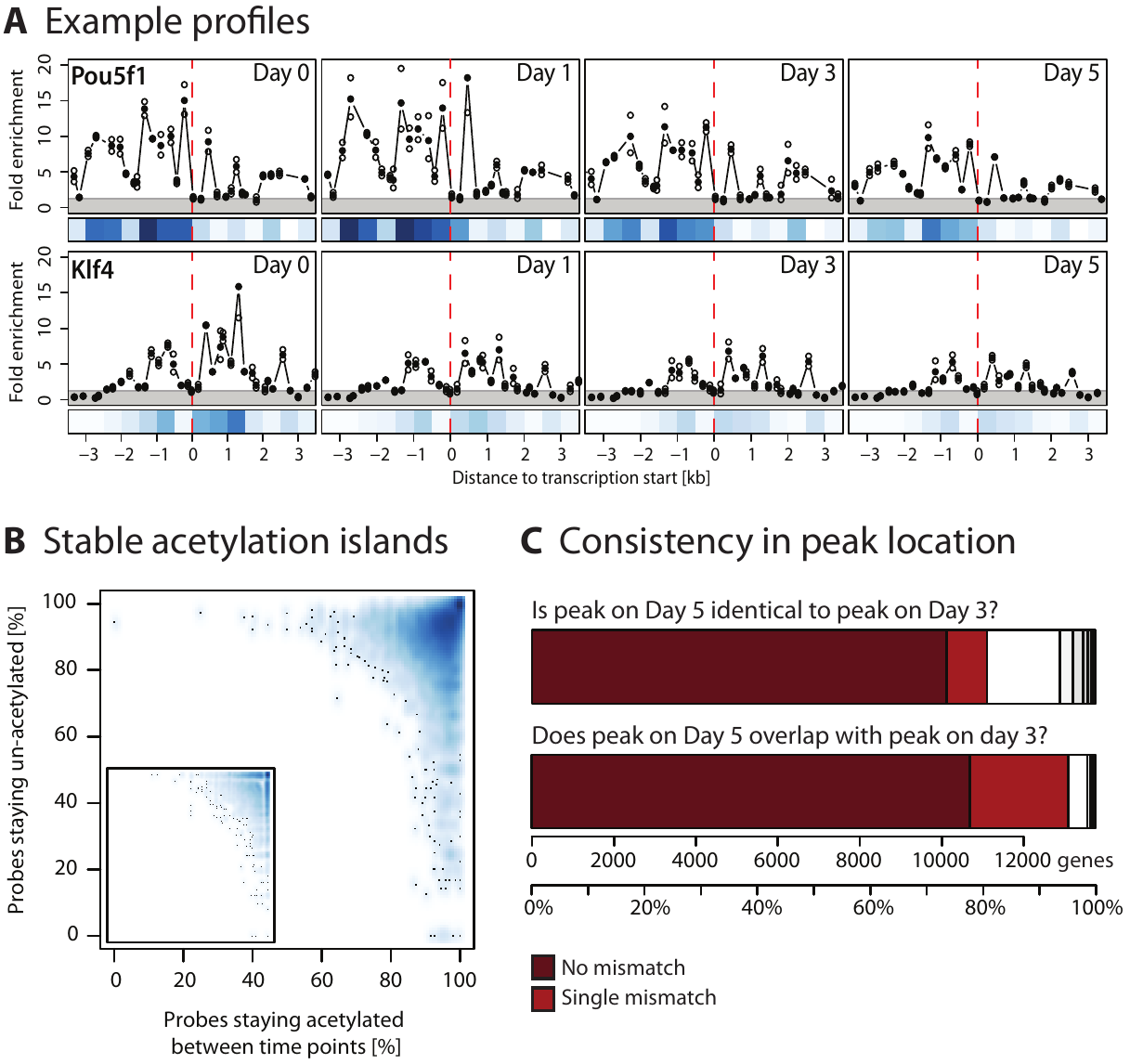}

\caption{\label{fig1}\textbf{Acetylation profiles over time} \textbf{A} Histone acetylation profiles of Pou5f1/Oct4 and Klf4 before Nanog-knockdown (day 0) and on days 1, 3, and 5 afterwards. All plots are centered at the transcription start site (TSS; red dashed line). The gray area shows the background signal, circles indicate replicate measurements, dots averages. The blue heatmap underneath each plot shows quantitative data averaged over .5kb intervals to make it comparable between genes with different numbers and positions of probes.  \textbf{B} To test for evidence of location changes, we counted probes as 'acetylated' if they were above the noise (gray area in panel A). The smoothed scatterplot shows for each gene the percentages of probes staying acetylated (x-axis) or un-acetylated (y-axis) over time. The mass of the distribution lies in the upper right corner indicating high stability of acetylation islands. This is independent of particular gene sets of days as the inlay exemplifies by plotting only the changes between day 3 and 5 for genes differential on day 5. \textbf{C} We defined a peak in the acetylation profile as the smallest region covering 30\% of the total signal.  Peaks stay very stable over time. The plot shows that for example between days 3 and 5 ca. 70\% of peaks are at exactly the same position and for almost 80\% of peaks the location on day 5 overlapped the location on day 3 completely.  If we allow one mismatch between peak locations the numbers go up to 80\% and 95\% respectively.}
\end{figure}


\begin{figure}[h!]
\includegraphics{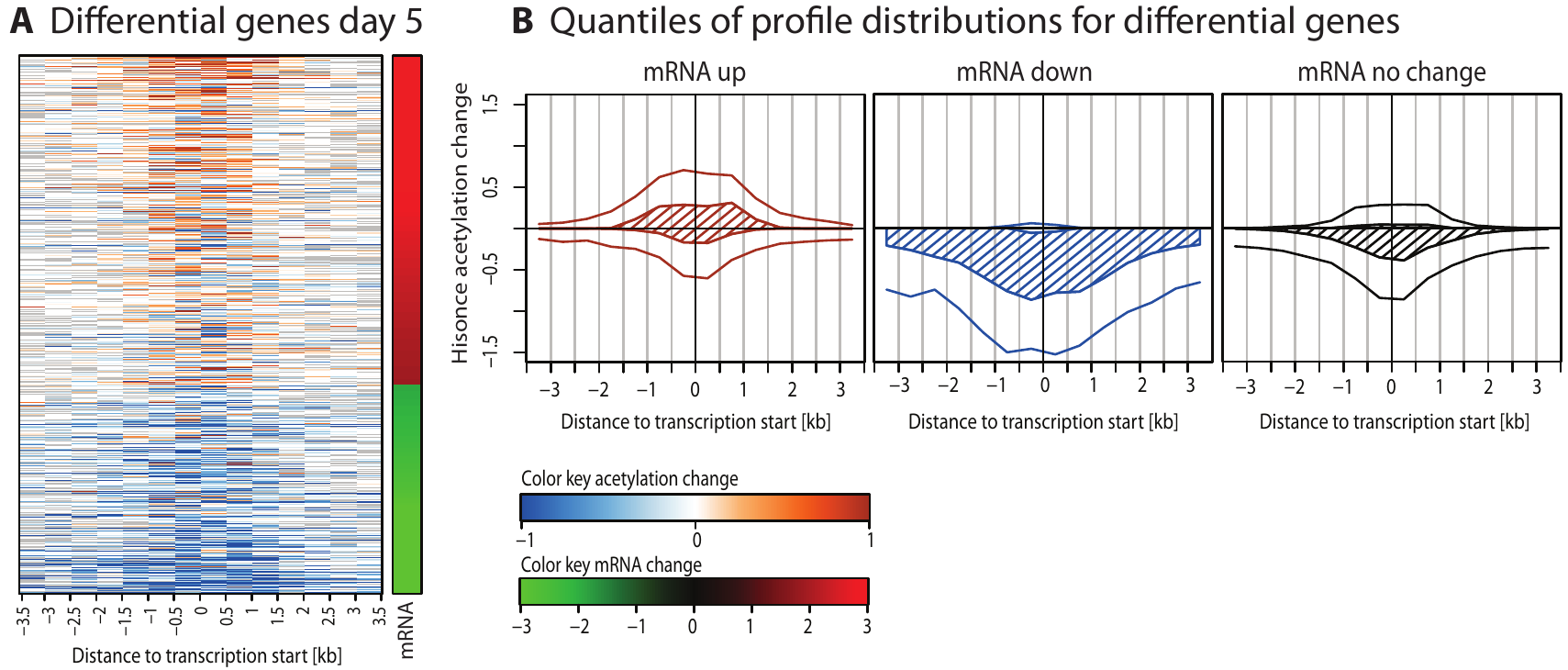}

\caption{\label{fig2}\textbf{Acetylation profiles for differential genes} 
\textbf{A} A heatmap of acetylation changes between days 0 and 5 for all genes with significantly differential mRNA levels on day 5. Transcriptionally up-regulated genes show an increase in acetylation signal, while down-regulated genes show a decrease. \textbf{B} Visualization of the distributions of changes in acetylation signal from day 0 to day 5 for genes transcriptionally up-regulated, down-regulated or non-changing on day 5. Each plot shows four lines corresponding to the 10\%, 25\%, 75\%, and 90\% quantiles of the distributions in each bin. The hatched area  emphasizes the inter-quartile range between the  25\% and 75\% quantile. Up-regulated genes show elevated acetylation levels close to TSS, while down-regulated genes show a broad decrease in acetylation across several kb around TSS.}
\end{figure}


\begin{figure}[h!]
\includegraphics{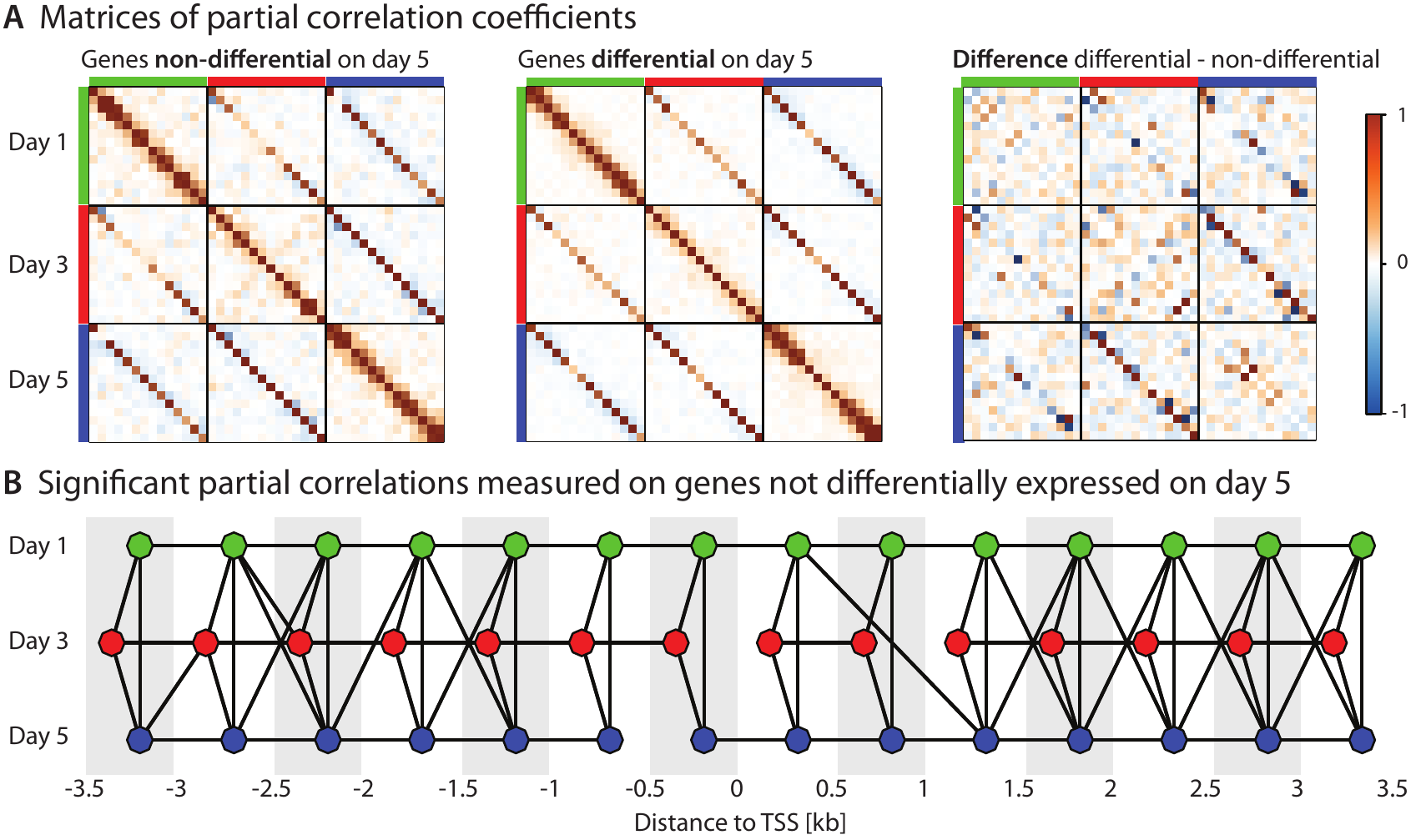}

\caption{\label{fig3}\textbf{Partial correlation analysis of acetylation profiles:} We analyzed spatial and temporal dependencies between regions around TSS by partial correlation coefficients. \textbf{A} Matrices of partial correlation coefficients for histone acetylation profiles on days 1 (green), 3 (red) and 5 (blue) computed on non-differential genes only (left) and differential genes only (middle). The right matrix shows the difference of the other two. \textbf{B} A graph representation of significant partial correlations (multiple testing corrected $p$-value $<0.05$). We show the graph computed on non-differential genes only. Partial correlations on differential genes are very similar, as panel A shows, but since there are many more non-differential than differential genes we gain in power to detect significant correlations. We find that spatial and temporal relationships are largely preserved in the partial correlation structure.  However, regions closer to TSS [$\leq$1.5kb] are less densely connected than the regions further away and in particular show gaps at positions right next to TSS on days 3 and 5. }
\end{figure}


\begin{figure}[h!]
\includegraphics[width=\textwidth]{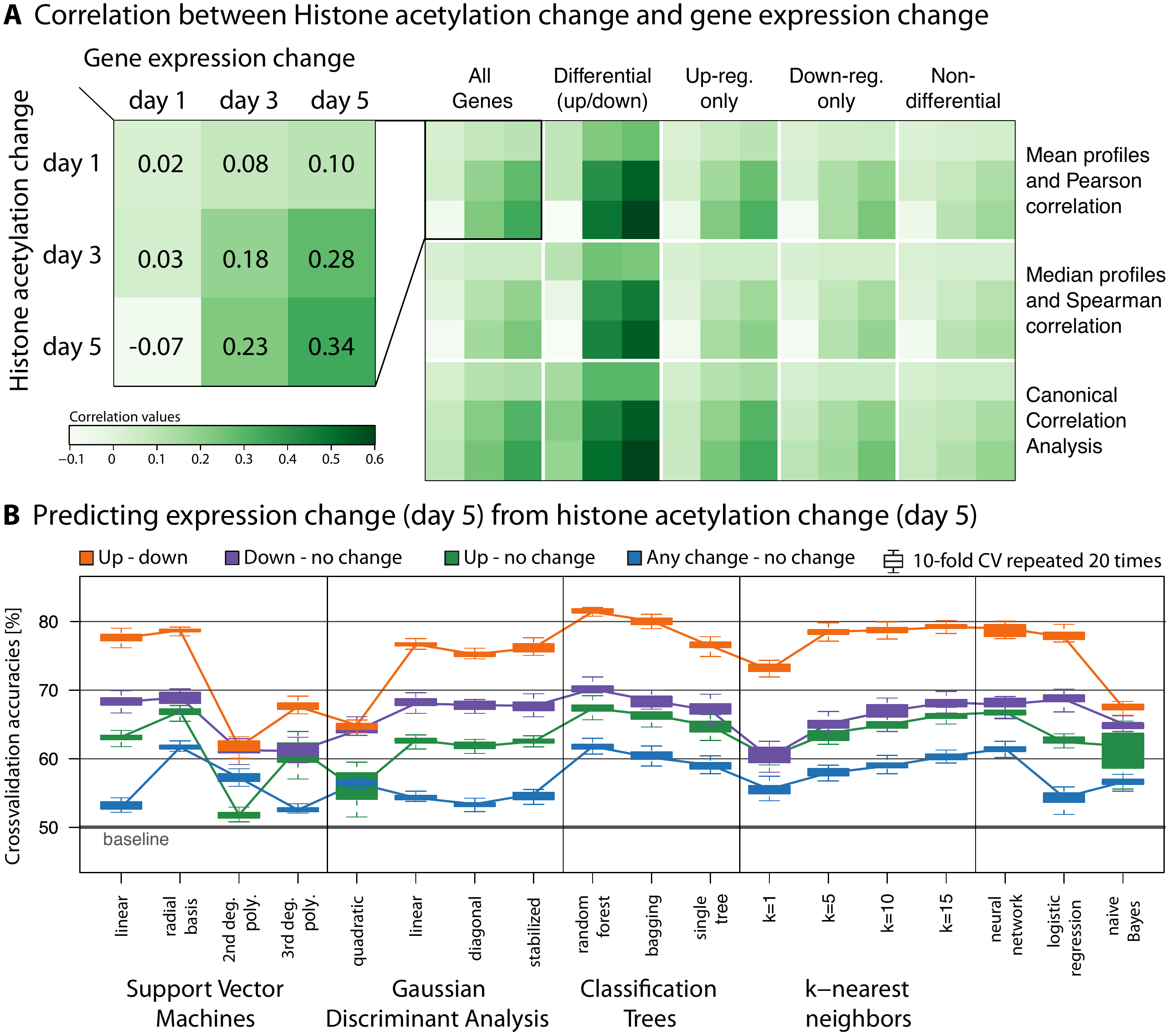}

\caption{\label{fig4}\textbf{Predictive power of acetylation changes for gene expression changes:}  \textbf{A} The left matrix shows the correlation between genome-wide mean acetylation changes and gene expression changes using Pearson correlation. Correlation values are small, but highly significant (see discussion in the main text). The right matrix shows correlation results when using other gene sets defined by differential expression on day 5 (columns of the matrix) or other measures of correlation (rows of the matrix).  \textbf{B}  Cross-validation results for a wide array of statistical classifiers predicting gene expression change from histone acetylation change. For each classifier four boxplots show the results of 10-fold cross-validation repeated 20 times sampling balanced data sets. The color of the boxplots corresponds to one of four classification problems:  Up- versus down-regulation (Orange), Down-regulation versus no-change (Purple), Up-regulation versus no-change (Green) and any change (up or down) versus no change (Blue). }
\end{figure}


\begin{figure}[h!]
\includegraphics{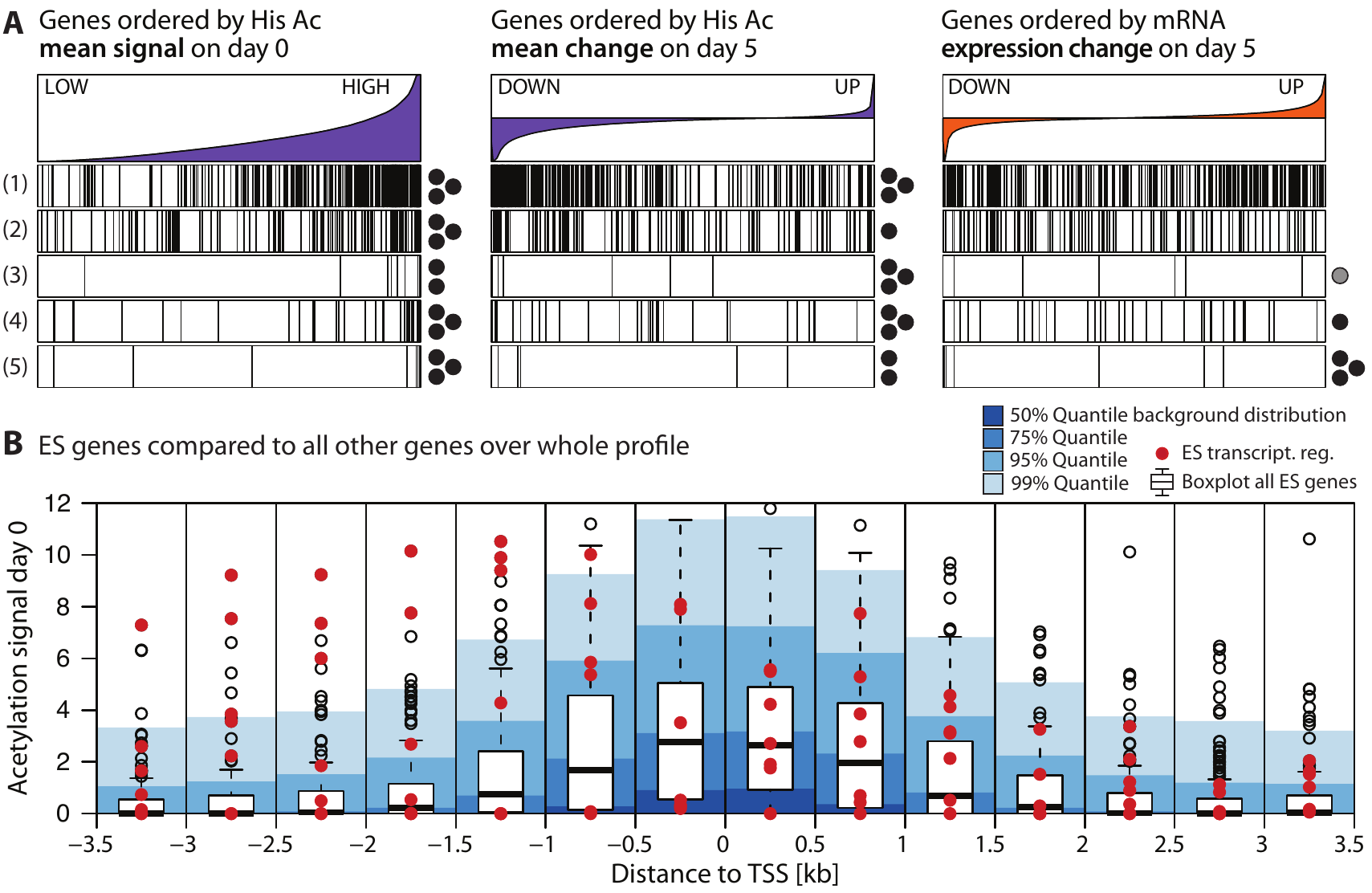}

\caption{\label{fig5}\textbf{ESC genes show distinct histone acetylation patterns:} We compare five sets of ESC specific genes to all other genes in terms of their histone acetylation and gene expression changes: (1) members of the PluriNet \cite{Mueller2008}; (2) hits of a recent RNAi screen \cite{Hu2009}; (3) gene ontology term GO:0019827 `stem cell maintenance'; (4) members of an ESC-specific protein-interaction network \cite{Wang2006};  (5) key transcriptional regulators \cite{Ivanova2006}.  \textbf{A} All genes are ordered by their mean acetylation signal on day 0, their acetylation change on day 5 and their expression change on day 5. The positions of the five ES specific gene sets in this ordering are then indicated by bars. The dots and circles indicate statistical significance of observed trends evaluted by GSEA: three dots for $p\leq10^{-4}$, two for $p\leq10^{-3}$ and one for $p\leq10^{-2}$, while a circle represents $p\leq0.05$.  \textbf{B} Here we compare ES genes to all others over the whole acetylation profile. The blue areas indicate quantiles of the genome-wide distribution of acetylation signal. The ES specific gene sets (white boxplots) show overall very high acetylation levels, in particular the transcriptional regulators (red dots) show surprisingly high histone acetylation levels before TSS.}
\end{figure}


\end{document}